\documentstyle[psfig]{mn}
\input{psfig.sty}

\newif\ifAMStwofonts

\title[Quiescent times in $\gamma$-ray bursts]
        {Quiescent times in $\gamma$-ray bursts: I. An observed correlation
        between the durations of subsequent emission episodes}
\author[Ramirez-Ruiz \& Merloni ]
        {Enrico Ramirez-Ruiz \&  Andrea Merloni
\\Institute of Astronomy, Madingley Road, Cambridge, CB3 0HA}

\date{}

\begin{document}

\maketitle

\label{firstpage}

\begin{abstract}

Although more than 2000 astronomical $\gamma$-ray bursts (GRBs) have
been detected, the precise progenitor responsible for these events is
unknown. The temporal phenomenology observed in GRBs can significantly
constrain the different models. Here we analyse the time histories
of a sample of bright, long
GRBs, searching for the ones exhibiting relatively long (more than 5 per cent of the total burst duration) {\it quiescent
times}, defined as the intervals between adjacent episodes of emission
during which the $\gamma$-rays count rate drops to the background
level. 
We find a quantitative relation between the
duration of an emission episode and the quiescent time elapsed since
the previous episode. We suggest here that the mechanism
responsible for the extraction and the dissipation of energy has to
take place in a {\it meta-stable} configuration, such that the longer
the accumulation 
period, the higher is the stored energy 
available for the next emission episode.            
\end{abstract}


\section{Introduction}
The study of $\gamma$-ray bursts (GRBs) has undergone a revolution since the
the first fading sources at X-ray \cite{costa97}, optical \cite{van97}
and radio \cite{frail97}
wavelengths were discovered, making
them the most powerful photon-emitters known in the Universe. Although
much attention has been devoted to the late afterglow emission since
then,  the prompt $\gamma$-ray
emission still has to be understood. Many issues remain unsolved, regarding the nature of the central engine, the
different scenarios giving rise to this prompt emission, and the
radiation mechanisms.

The data collected by the {\it Burst and Transient Source Experiment},
{\it BATSE}, have provided us with an unprecedented wealth of information. 
Nonetheless, gamma-ray bursts are so complicated and diverse in the time domain that, at first sight, their behaviour
obeys no simple rule. 

One of the main issues that has
remained largely unexplained is what determines the characteristic
durations of the bursts, that typically range 
between $10^{-2}$ and $10^{3}$ seconds\footnote{The definition of the
burst duration is not unique. The {\it BATSE} team characterizes it
using $T_{90}$ ($T_{50}$): the time needed to accumulate from 5\% to
95\% (from 25\% to 75\%) of the counts in the 50-300 keV
band \cite{meegan96}. For the purpose of our analysis, we used
$T_{90}$ as a measure of burst duration.}. 
Observationally \cite{kou93} the short ($\la 2
s$) and long ($\ga 2 s$) bursts appear to represent two distinct
subclasses. An early proposal made by Katz \& Canel (1996) to explain the bimodal distribution of durations was
that accretion-induced collapse of a white dwarf into a NS plus debris
might be a candidate for the long bursts, while NS-NS mergers could provide the
short ones. However, it is at present unclear which, if any, of these
progenitors is responsible for the bulk of GRBs. 

Besides this apparent bimodality, $\gamma$-ray burst temporal profiles
are enormously varied. Many  bursts have a highly variable temporal profile
with a variability time scale that is significantly shorter than the overall
duration, while in a minority of them there is only one peak, with no
substructure. Furthermore, long $\gamma$-ray bursts' time
histories often show multiple episodes of emission, separated by
background intervals,  or {\it quiescent
times}, of variable durations. 
In other words, it seems that the emission can turn off 
to a very low level and then turn on again. 
This observed property can provide an
interesting clue to the nature of $\gamma$-ray bursts. At present, it
is unclear if these separated emission episodes are consequences of the
same physical process (e.g. internal or external shocks), and if the time 
separation is due to some intrinsic property of the central source or 
of its environment. 

The purpose
of this paper is to determine the properties of quiescent times in
long $\gamma$-ray bursts observations. In an accompanying paper
(Ramirez-Ruiz, Merloni \& Rees  2000, hereafter Paper II) we study, within
the framework of the internal-external shock model, the various
possible mechanisms that can give rise to quiescent times in the
observed $\gamma$-ray light-curves.
    
\section{Observations}

\subsection{Procedure}

A visual inspection of the {\it BATSE} catalogue of multi-peaked time histories
reveals that in some of them the count rate drops to the background level 
in between two adjacent episodes of higher emission intensity. Our aim
is to characterize and measure these episodes of quiescence. 

In an earlier report, Koshut et
al. (1995) found  that about 3\% of the 
bursts show signs of {\it precursor} activity, with a peak intensity 
lower than the main burst,  separated from the remaining emission  by
a background interval that is at least as long as the rest of the
burst. {\it BATSE} burst 2156, shown in the upper panel of Figure 1, falls into
this category. However, the above definition singles out a data set
with rather extreme properties.  Several time profiles show
periods of quiescence 
without having precursor activity. See, as an example, BATSE burst
3067, which is shown in the lower panel of Figure 1. For this reason,
in order to study the periods of quiet emission, we adopted
 more general selection criteria. Here, we allowed multiple pairs of
successive events separated by a quiescent time 
within a single burst and  we did not  impose
requirements on the relative intensities or on the
time interval separating any two emission episodes. The periods
of emission occurring before and after the quiescent time are 
therefore referred to as {\it
prequiet} and {\it afterquiet} time, respectively.

\begin{figure}
{\vfil 
\psfig{figure=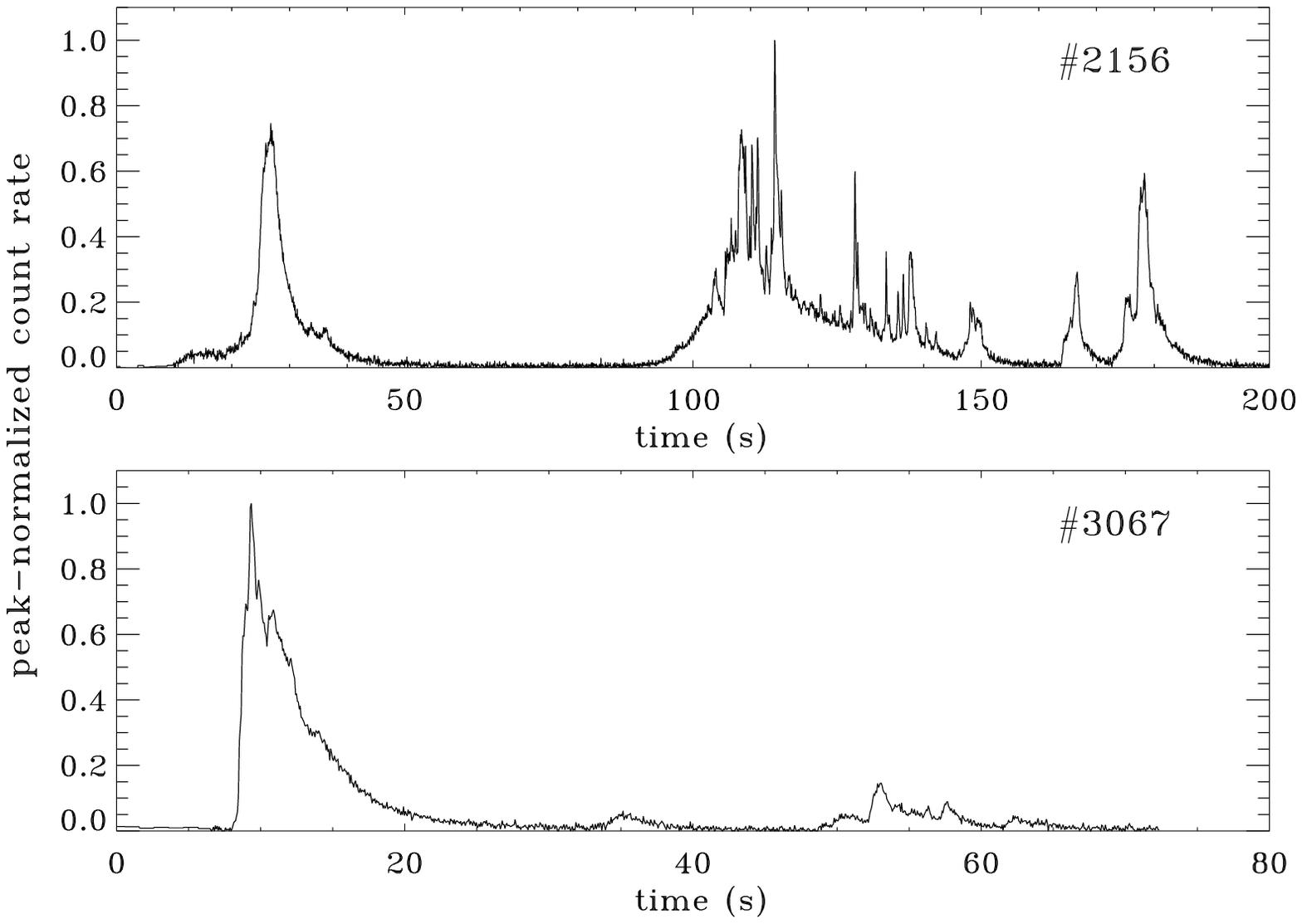,angle=0,height=75mm,width=0.5\textwidth}
\caption{Time profiles observed with BATSE (at energies $>$ 20keV) that
contain periods of quiet emission. Upper panel: BATSE burst
\#2156 have a very strong main burst after a long quiet emission period. 
Lower panel: BATSE burst \#3067 have a very strong main burst before
the count rate drops to the background level.}
\vfil}
\label{fig1}
\end{figure}

For the purpose of our analysis,
we have used all 94 bursts from the 4B {\it BATSE} catalog longer than
5s ($T_{90} > 5s$) and brighter than 5 photons$^{-1}$ cm$^{-2}$
({\it BATSE} peak photon flux in the 256 ms time-scale). 
We have used the {\it BATSE}
64 ms four-channel data (i.e., from 25 to $\sim$ 800 keV). A background model
was created for each energy channel by fitting user-defined background
intervals with a second-degree polynomial and interpolating this fit
across the source interval. We subtracted the background model from
the observed count rates in each energy channel.
This then gave us the total source count rates as a function
of time. In order to be selected, a burst must have at least one quiescent period in its
time history. The selection was accomplished by calculating the total number of
counts, over the entire energy range, recorded in a temporal window sliding
along the time axis. The width of the window was set to 5\% of the duration
of the burst. Thus, the width of the window varied
 from burst to burst, 
allowing us to avoid bias against quiescent times with duration
less than some arbitrary window width. We define a drop to the
background level whenever the number of net counts in a window is smaller than
the 2$\sigma$ level in the corresponding background counts window. We have used
this absolute test, rather than a relative limit (e.g., a fixed fraction
 of the highest
number of counts in any given window), to try to avoid the
possibility of the existence of source emission below our detection level,
during any time interval in any burst.

We found that $\sim$ 15\% of the analysed long and bright bursts
contain at least one quiescent interval in their time history (of duration $\ga
0.05 T_{90}$) and $\sim$ 25\%
of these have more than one.  
Figure 2 illustrates the distribution of durations for, respectively, all bursts in the sample (dotted line),
the subset of those that contain at least one quiescent period (dashed line), and
the bursts with two or more quiet emission intervals
(solid line).

\begin{figure}
{\vfil 
\psfig{figure=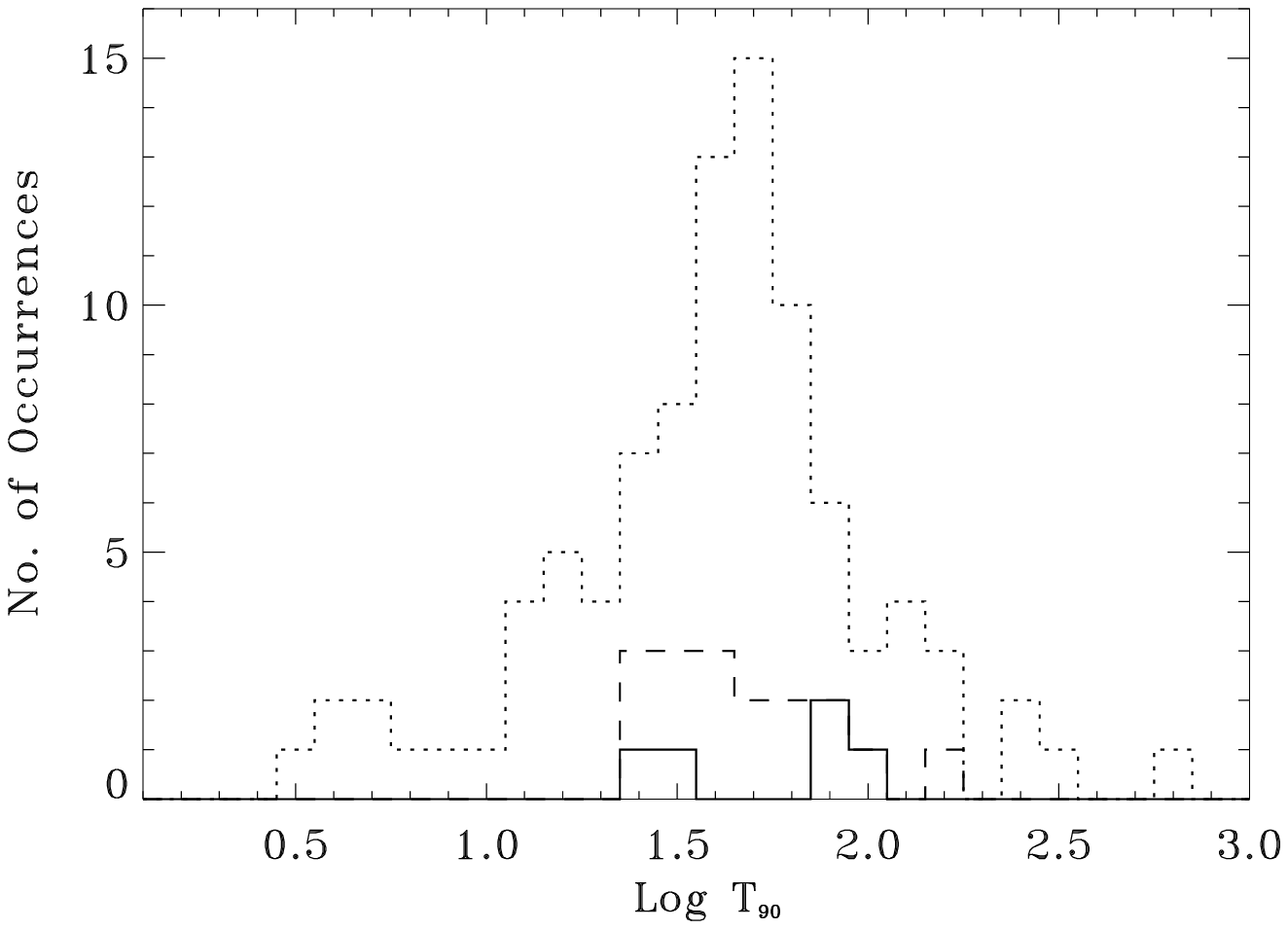,angle=0,height=80mm,width=0.5\textwidth}
\caption{Distribution of GRB durations. The dotted line is the
distribution of all bright bursts that were longer than
5s and brighter than 5 photons$^{-1}$ cm$^{-2}$, the dashed one
represents the bursts which show at least one quiescent time and the
solid one the bursts with more than one quiescent interval. }
\vfil}
\label{fig2}
\end{figure}

Figure 2 shows that the distribution of durations for the
subset of bursts that contained quiet emission periods is consistent
with the sample distribution. We find no significant evidence that the
presence of quiescent times is preferentially found in longer or shorter bursts
within our sample. Some limitations are necessarily inherent in our
approach and data selection: the conclusions we reach are based on
measurements of a subset of the bursts detected with {\it BATSE}; we analyse
relatively bright bursts with durations greater than $\sim$ 5 s,
recorded with 64 ms temporal resolution and four channel spectral
resolution. Analysis of quiescent times in shorter bursts will be more
difficult due to the fact that few short bursts are
brighter than 5 photons$^{-1}$ cm$^{-2}$. The selection of a high
brightness sample is appropriate in order to avoid the systematic
effects that might change the observed time histories with different
statistics. The time histories of dim events would be more randomized
by fluctuations than the time histories of bright bursts.
Using other GRB samples with a high signal to noise level ($\ga 3$ 
photons$^{-1}$ cm$^{-2}$) gives similar results.

\subsection{Results: correlations between emission properties}

We have searched for correlations between the temporal
properties of the different emission periods. Any correlation will
provide strong constraints on various burst models.
In particular, we have investigated the dependence of the duration of the
quiescence period on the {\it afterquiet} and the {\it prequiet} burst durations. It is worth stressing that, by comparing time intervals within each burst,
we eliminate the distance dependence (or time dilation effects) that
would arise if we compared, for example, total number of counts of an emission
episode with the duration of the quiescence period. We find no evidence for a
correlation between quiescent times  and prequiet burst
times, as seen in Figure 3a. However, a strong
one-to-one correspondence seems to exist between quiescent times
and afterquiet burst durations, at about  4$\sigma$ confidence
level ($r_{s}$ $\sim$ 0.89). Stated otherwise, we found that, in our
time histories sample, the longer the quiescent time the longer the
duration of the following emission period, as shown in Figure 3b.

\begin{figure}
\vbox to120mm
{\vfil 
\psfig{figure=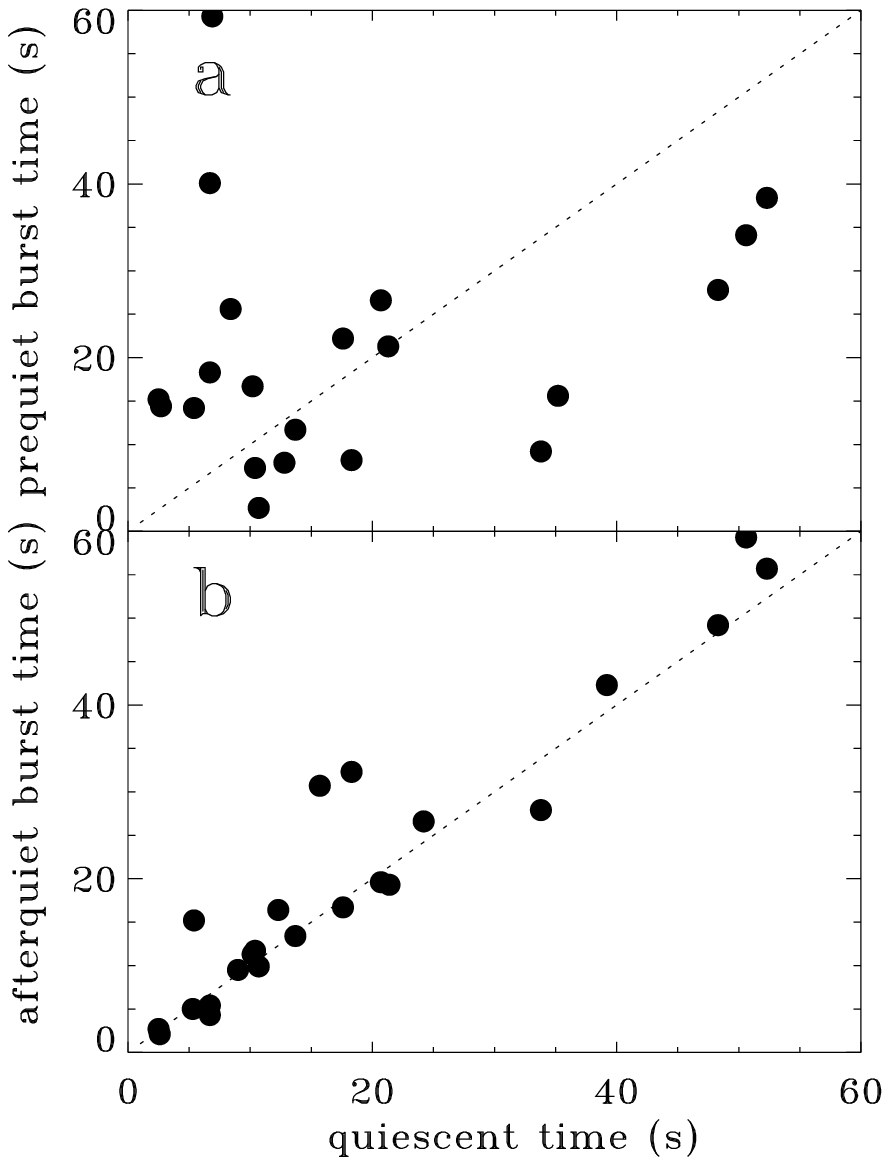,angle=0,height=90mm,width=150mm}
\caption{Correlations between the temporal properties of the different
emission periods. In panel (a) we plot the duration of an emission
episode ({\it prequiet burst time}) against the duration of the
following quiescent time. No clear correlation is found in this case.  
In panel (b) we plot the duration of an emission episode ({\it
afterquiet burst time}) against the duration of the previous quiescent
time. There is a clear trend: the longer the quiescent time, the
longer the duration of the following emission period.  
}
\vfil}
\label{fig3}
\end{figure}

Lochner (1992) studied 
the relation between successive emission episodes in multiple episode bursts
observed with the Pioneer Venus Orbiter ({\it PVO}) gamma-ray burst
detector. Unlike the work presented here, his data set was only a
collection of multiple event bursts and did not result from a systematic
search throughout the entire {\it PVO} database. 
Lochner (1992) reported a strong
correlation ($r_{s} \sim 0.73$) between the duration of an event and the
duration of the subsequent event. 

In the study of precursor activity \cite{koshut95}, 
a similar trend was found. However, in their analysis such correlation
is simply a consequence of the definition of precursor activity, since
the authors required that the two sub-bursts were separated by a background
interval long at least as the remaining emission episode. 

Furthermore, Lochner (1992) reported also a moderate correlation between 
the hardness of an event and the hardness of the following event. 
We found no evidence for such a property in our sample.

The existence of a correlation 
between the duration of the quiescent time and the {\it total energy} of the
afterquiet burst would of course reveal very interesting 
properties of GRB sources. 
However, to characterize the total energy of a
burst, we would need information about the distance of the
event\footnote{At the moment, there is only one {\it BATSE} event (7560) 
with  periods of quiet emission in its light-curve and for which a redshift measurement has been done (GRB 990510).}. 
Without this information, looking for any correlation  between the total
number of counts (and hence energy) of the afterquiet emission 
and the duration of the quiet period in different events will be misleading.

Nonetheless, 
one could expect a correlation between the duration of
an emission episode and the total energy radiated. If this is the case, the correlation between times we found should simply reflect 
a correlation between the burst strength and the time elapsed
since the previous emission episode. 
It  will then be indicative of an accumulation of
fuel, similar to what observed in the galactic
superluminal jet source GRS 1915+105 \cite{Bell97}. On the other hand,
a correlation between the burst strength and 
the time until the next burst, indicative of a relaxation oscillator
behaviour, is ruled out by the observations.

\section{Discussion}

Here we present a few simple considerations that can be drawn from the
hypothesis that the gaps in the $\gamma$-ray light curves of GRBs are a
consequence of a central engine which actually goes dormant for a
period of time comparable to the duration of the gaps.
To falsify this hypothesis clearly requires a thorough scrutiny of all
the alternative possibilities. These possibilities will be addressed
in detail in Paper II. In particular, in the framework of the internal
shock model, we show that there are 
realistic assumptions that produce a long
quiescent interval in the light-curve of a gamma ray burst, without
having to postulate that the central source itself turns off  for a
comparably long time.
  
All the bursts analysed here are long and structured. Clearly in these
cases the central engine has to be active for a  period extremely
long compared to the typical dynamical time-scale
($\sim$ milliseconds) for stellar-mass compact objects. 
Thus, the central engine has to evolve into a configuration which
is stable enough 
to survive the violent gravitational instabilities associated with the
merging/collapse of compact objects, while still keeping enough binding energy to power the burst. 
A thick torus (or an advective, optically thick accretion disk)
accreting at a rate of $0.01$ to $10$ $M_{\odot} {\rm yr}^{-1}$ (Popham,
Woosley \& 
Fryer 1999, hereafter PWF), is a key ingredient for achieving this. 
The system also needs to be highly unstable to produce the extremely
varied light curves we observe \cite{Stern99}, and this requirement is
even stronger for the bursts exhibiting quiescent times. 
  
Of the two more popular mechanisms that have been proposed to explain the GRBs
energy release (neutrino annihilation and conversion of Poynting flux
into a magnetized wind), the first seems unable to produce the longer,
more energetic and variable bursts (PWF; Rees 1999).  
As PWF have shown, the efficiency
of such a process is highly variable and extremely sensitive to the
accretion rate: higher accretion rates lead to higher efficiencies. They
conclude that neutrino annihilation in hyper-accreting black hole
systems can explain bursts of energy up to $10^{52}$ergs. 

The other popular scenario, the conversion of
Poynting flux into a magnetized wind, requires, in order to liberate 
the observed amount of energy, a magnetic field $\ga
10^{15}$ G (Rees 1999, and references therein). 
Such a high field is
not unphysical. First, because it is about two orders of
magnitude smaller than the virial limit; second, because it has
possibly been observed in some 
peculiar neutron stars (the so called {\it magnetars}; see e.g. 
Duncan \& Thompson 1992; Kouveliotou et al. 1998, and references therein). 

 A neutron torus, with its huge amount of differential
rotation, is a natural site for the onset of a dynamo process that
winds up the field to the required intensity \cite{KR98}, provided that
a fluid element is able to complete a sufficient number of orbits around the
hole. This number, in turn, depends on the viscosity inside the disk,
$N \sim \alpha^{-1}$, where $\alpha$ is the standard Shakura-Sunyaev viscosity. 

The actual properties of the expected variability depend on the details
of the configuration of the disc corona generated by the magnetic
field, which is removed from the disc interior via turbulent flux buoyancy
\cite{Araya99}.  This is a very complicated (and almost unaddressed;
but see Tout \& Pringle 1996) issue, on which some insight can come
from the observations. 
 
The tentative correlation found between the duration of an emission
episode in a multi-peaked burst and the duration of the preceding
quiescent time hints at the following general scenario. 

The system builds up its energy (via an MHD instability driven dynamo,
for example) and reaches a near critical, or {\it meta-stable}
state. Any local instability can, by definition,
cause a rapid dissipation of all the stored energy through an avalanche of
dissipation events. The system will tend
to return to a more stable configuration, characterized by a certain
threshold energy $E_0$, or a sub-critical coronal magnetic field
configuration. The source then becomes quiescent. If the lifetime of
the accreting torus is long enough  and
depending on the rate at which the energy is actually extracted from
the disc (or from the black hole) and deposited into the external magnetic
field, the system can undergo another episode of strong emission. 
As we can assume $E_0$
to be fixed by the geometry and by the physical parameters of the 
black hole--accretion disc system, the longer the
quiescent time, the higher will be the stored energy above the
threshold available for the next episode. 
Such a situation will give rise to the observed correlation. 

This is a mechanism different from any relaxation oscillator, in which
the threshold energy is an {\it upper} limit for the system. As soon
as the system reaches such a limit it is forced to release energy: in this
case, the larger 
the amount of energy released, the longer will be the time needed to
reach the threshold again, and we would obtain a correlation between
the burst time and the prequiet time, contrary to what observed.

It is also important to note that the threshold energy $E_0$
above which the system is in a meta-stable state is not directly
related to the intensity of the seed magnetic field inside the neutron
torus. It is instead most likely related to the intensity and the
topological configuration of the amplified field which emerges from
the torus. 

Finally, we would like to point out the suggestive analogy with the 
microquasar GRS 1915+105, a galactic black hole candidate which 
exhibits a strikingly similar (even quantitatively) correlation between 
the duration of a quiescent time and that of the following burst 
(Belloni et al. 1997). 
This source is believed to accrete at a rate very close to 
(or maybe higher than) the Eddington limit, which is also the case for 
the configuration suggested for gamma-ray bursters discussed above.

\section{Conclusions}

The very existence of quiescent times in GRB light curves poses severe
restrictions on the emission models (see Fenimore \& Ramirez-Ruiz 1999;
Paper II) and, possibly, gives
direct insight into the 
dynamical properties of the hidden central engine. 

Apart from the energy requirements that any viable model has to fulfill,
we envisage that the observed temporal structure will bring
fundamental information to the understanding of GRBs progenitors. 

We have systematically analysed a sample of bright, long gamma-ray
bursts searching for the ones exhibiting relatively long quiescent
times (more than 5\% of the total burst duration). These amount to
$\sim 15\%$ of the sample. 
We have found an interesting correlation between the duration of the 
quiescent interval and the duration of the following emission
episode, while no correlation has been found between the quiescent
time and the duration of the previous emission episode. 

We suggest here that, in the hypothesis that the gaps in the observed
light-curves directly reflect a period of inactivity of the central
source, the mechanism 
responsible for extracting and dissipating the energy in the high
Lorentz factor ejecta has to develop a {\it meta-stable} configuration. That
is, a configuration in which a local instability can abruptly drain
the system of all the stored energy, probably via a cascade of
correlated smaller scale events.

\section*{Acknowledgments}
We thank M. Rees, P. Natarajan, P. Madau, A. Fabian and G. Morris for useful comments and suggestions. ERR acknowledges support from
CONACYT, SEP and the ORS foundation. AM thanks PPARC and the TMR network `Accretion onto black holes, compact stars and protostars', funded by the European Commission under contract number ERBFMRX-CT98-0195, for support.

\bsp

\label{lastpage}


\begin{thebibliography}{99}


\bibitem[\protect\citename{Araya-G\'ochez }1999]{Araya99}
Araya-G\`ochez, R. A., 1999, astro-ph/9912324.


\bibitem[\protect\citename{Belloni et al. }1997]{Bell97}
Belloni T., M\'endez, M., King, A. R., van der Klis, M. \& van Paradijs, J., 1997, ApJ, 488, L109.

\bibitem[\protect\citename{Costa et al. }1997]{costa97}
Costa, E., et al. 1997, Nature, 387, 783.

\bibitem[\protect\citename{Duncan \& Thompson }1992]{DC92}
Duncan, R. C. \& Thompson, C. 1992, ApJ, 392, L9.

\bibitem[\protect\citename{Fenimore \& Ramirez-Ruiz }1999]{fr99}
Fenimore, E. E. \&  Ramirez-Ruiz, E., 1999, PASP Conf. Proc. Gamma-Ray
Bursts: The First Three Minutes, astro-ph/9906125.

\bibitem[\protect\citename{Frail et al. }1997]{frail97}
Frail, D., Kulkarni, S. R., Nicastro, L., Feroci, M. \& Taylor, G. B.,
1997, Nature, 389, 261.

\bibitem[\protect\citename{Katz \& Canel }1996]{katz96}
Katz, J. \& Canel, L. M., 1996, ApJ, 471, 915.

\bibitem[\protect\citename{Klu\'zniak \& Ruderman }1998]{KR98}
Klu\'zniak, W. \& Ruderman, M., 1998, ApJ, 505, L113.

\bibitem[\protect\citename{Kouveliotou et al. }1993]{kou93}
Kouveliotou, C., et al., 1993, ApJ, 413, L101.

\bibitem[\protect\citename{Kouveliotou et al. }1998]{Kou98}
Kouveliotou, C., et al., 1998, Nature, 393, 235.

\bibitem[\protect\citename{Koshut et al. }1995]{koshut95}
Koshut, T. M., et al., 1995, ApJ, 452, 145.

\bibitem[\protect\citename{Lochner }1992]{lochner92}
Lochner, J. C., 1992, in AIP Conf. Proc. 265, Huntsville Gamma-Ray
Burst Workshop, ed. W. S. Paciesas \& G. J. Fishman (New
York:AIP),289.

\bibitem[\protect\citename{Meegan et al. }1996]{meegan96}
Meegan, C. A., et al., 1996, ApJS, 106, 65.

\bibitem[\protect\citename{Popham, Woosley \& Fryer }1999]{PWF99}
Popham, R., Woosley, S. E. \& Fryer, C., 1999, ApJ, 518, 356. (PWF)

\bibitem[\protect\citename{Ramirez-Ruiz et al. }2000]{err200}
Ramirez-Ruiz, E., Merloni, A. \& Rees, M. J., MNRAS, submitted. (Paper II)

\bibitem[\protect\citename{Rees }1999]{Rees99}
Rees, M. J., 1999, A\&AS, 138, 491.


\bibitem[\protect\citename{Stern }1999]{Stern99}
Stern, B., 1999, ASP Conf. Series, Vol. 190, 31.

\bibitem[\protect\citename{Tout \& Pringle }1996]{TP96}
Tout, C. A. \& Pringle, J. E., 1996, MNRAS, 281, 219.

\bibitem[\protect\citename{van Paradijs et al. }1997]{van97}
van Paradijs, J. et al., 1997, Nature, 386, 686. 

\end{thebibliography}
\end{document}